\documentclass[preprint,aps]{revtex4-1}
\usepackage{soul}
\usepackage{graphicx}
\usepackage{epstopdf}
\usepackage{bm}
\usepackage{upgreek}
\usepackage{amsmath}
\usepackage{lipsum}
\usepackage{color,xcolor}

 \usepackage[colorlinks,linkcolor=blue,
            citecolor=blue,
            hyperindex,
            pdfstartview=FitH,
            plainpages=false]
            {hyperref}
\newcommand {\BSCCO}{Bi$_2$Sr$_2$CaCu$_2$O$_{8+x}$}
\newcommand {\BSCCOs}{Bi$_2$Sr$_2$CaCu$_2$O$_{8+x}$ }
\newcommand {\PbBSCCO}{(Bi$_{2-x}$Pb$_x$)Sr$_2$CaCu$_2$O$_{8+x}$}
\newcommand {\BSCCOsingle}{Bi$_2$Sr$_2$CuO$_{6+x}$}
\newcommand {\BSCCOsingles}{Bi$_2$Sr$_2$CuO$_{6+x}$ }

\definecolor{CYgray}{rgb}{0.4, 0.4, 0.4}
\definecolor{CYred}{rgb}{0.6, 0.13, 0.35}
\definecolor{CYgreen}{rgb}{0.2, 0.5, 0.25} 
\usepackage{lineno}

\begin{document}
\begin{sloppypar}

\title{Doping Dependence of Spin-Momentum Locking in Bismuth-Based High-Temperature Cuprate Superconductors}

\author{Hailan Luo$^{1,2}$, Kayla Currier$^{1,2}$,Chiu-Yun Lin$^{1,2}$, Kenneth Gotlieb$^{2,3}$, Ryo Mori$^{2}$, Hiroshi Eisaki$^{4}$, Alexei Fedorov$^{5}$, Zahid Hussain$^{5}$ and Alessandra Lanzara$^{1,2*}$
}

\affiliation{
\\$^{1}$Department of Physics, University of California, Berkeley, California 94720, USA
\\$^{2}$Materials Sciences Division, Lawrence Berkeley National Laboratory, Berkeley, California 94720, USA
\\$^{3}$Graduate Group in Applied Science \& Technology, University of California, Berkeley, California 94720, USA
\\$^{4}$Research Institute for Advanced Electronics and Photonics, National Institute of Advanced Industrial Science and Technology, Tsukuba, Ibaraki 305-8568, Japan
\\$^{5}$Advanced Light Source, Lawrence Berkeley National Laboratory, Berkeley, California 94720, USA
\\$^{*}$Corresponding author: alanzara@lbl.gov
}

\date {\today}

\vspace{8mm}

\begin{abstract}

Non-zero spin orbit coupling has been reported in several unconventional superconductors due to the absence of inversion symmetry breaking. This contrasts with cuprate superconductors, where such interaction has been neglected for a long time. The recent report of a non-trivial spin orbit coupling in overdoped Bi2212 cuprate superconductor, has re-opened an old debate on both the source and role of such interaction and its evolution throughout the superconducting dome. Using high-resolution spin- and angle-resolved photoemission spectroscopy, we reveal a momentum-dependent spin texture throughout the hole-doped side of the superconducting phase diagram for single- and double-layer bismuth-based cuprates. 
The universality of the reported effect among different dopings and the disappearance of spin polarization upon lead substitution, suggest a common source. We argue that local structural fluctuations of the CuO planes and the resulting charge imbalance may cause local inversion symmetry breaking and spin polarization, which might be crucial for understanding cuprates physics.
\end{abstract}

\maketitle
{\bf Introduction}

Symmetry and symmetry breaking are arguably among the most fundamental concepts in condensed matter physics and more generally are key to classifying and understanding a variety of natural phenomena \cite{Enquist1994}. The properties of materials can be linked to their various global symmetries, such as time and space inversion symmetry. When symmetry is broken in crystals, new patterns emerge and give rise to novel phases and properties \cite{Landau1937}. For instance, the removal of inversion symmetry can lead to coupling of the electron’s orbital motion with its spin, also known as spin-orbit coupling (SOC), eventually resulting in two spin-split electronic bands in momentum. 

The ramification of global inversion symmetry breaking in the context of superconductivity has been a matter of study for long time \cite{Bauer2012, Smidman2017}, with consequences on the pairing symmetry via singlet-triplet mixing, and with the general understanding that spin-orbit coupling mostly behaves as a symmetry breaking field to the superconducting state \cite{Gorkov2001, Frigeri2004}. However, the recent realization that breaking the local symmetry could also drive formation of novel exotic phases, even when the global symmetry is preserved, has opened a new paradigm in materials design. In the specific case of inversion symmetry, local symmetry breaking can still lift the spin degeneracy through an asymmetry of the wave function, while bearing a zero net spin polarization \cite{Zhang2014, Riley2014,Yuan2019}. In this case, the resulting spin-locking pattern is not necessarily that of the textbook's Rashba effect at $\Gamma$, but can occur across different valleys \cite{Bawden2016} or layers \cite{Yao2017}. 
 
The concept of local symmetry breaking becomes even more appealing in the context of superconductivity, where it has been proposed that it could play a key role in determining which competing phases are stabilized, in favoring of a specific symmetry of the superconducting order parameter, and ultimately even in enhancing the superconducting transition temperature \cite{Fischer2011}. 
Unconventional superconductors such as cuprates, pnictides, and even the recently reported 2D superconducting heterostructures are particularly attractive for studying the interplay between spin-orbit coupling and superconductivity, as often the local symmetry breaking is intrinsic to their nature  \cite{VVelasco2022SConradson, Cao2018, Bozovic2001, Reyren2007}. In the case of cuprates and pnictides for example, short-range structural distortions, rotations of the octahedra \cite{Bianconi1996, Billinge1994}, and staggered stacking between layers \cite{Sigrist2014}, are all possible sources of local symmetry breaking, with the latter being also shared among the majority of 2D superconducting heterostructures \cite{Du2021}. In each one of these cases, it is therefore natural to expect that the altered symmetry of the lattice, combined with the presence of SOC will have an impact on the order parameter \cite{Fischer2011}. The level of sophistication that has been reached in measuring the spin-resolved electronic structure of materials \cite{Okuda2017,Lin2021}, together with the recent report of a non-zero spin-momentum locking in a centrosymmetric cuprate \BSCCO (Bi2212) superconductor \cite{Gotlieb2018}, has opened a new frontier in this line of research.  However, after this first report confirmed by subsequent studies  \cite{Liu2023, Iwasawa2023}, there is still no confirmation whether this is a universal property of cuprates and how it evolves across the superconducting phase diagram. 

Here we use high resolution spin- and angle-resolved photoemission spectroscopy (spin ARPES) to study the doping and spin-dependent electronic structure of the single layer \BSCCOsingle (Bi2201),  double layer \BSCCO (Bi2212) and Pb-doped $\rm(Bi_{2-\it x}Pb_{\it x})Sr_2CaCu_2O_{8+{\it x}}$ (Pb-Bi2212). We reveal the persistence of an intrinsic and universal spin texture in the Bi-based hole-doped side of the cuprate phase diagram over the entire superconducting dome. We attribute such spin texture to a breaking of local inversion symmetry induced by the short range structural distortions of the CuO octahedra, and show how their reduction can lead to a suppression of the spin polarization. These findings reveal the imprint on the electronic structure by a combination of spin-orbit coupling and lattice symmetry, bringing forward the intriguing possibility that such interplay might constitute a missing ingredient in understanding the cuprate physics.
\vspace{3mm}

{\bf Results and Discussion}

\begin{figure*}[ht]\includegraphics[width=14 cm]{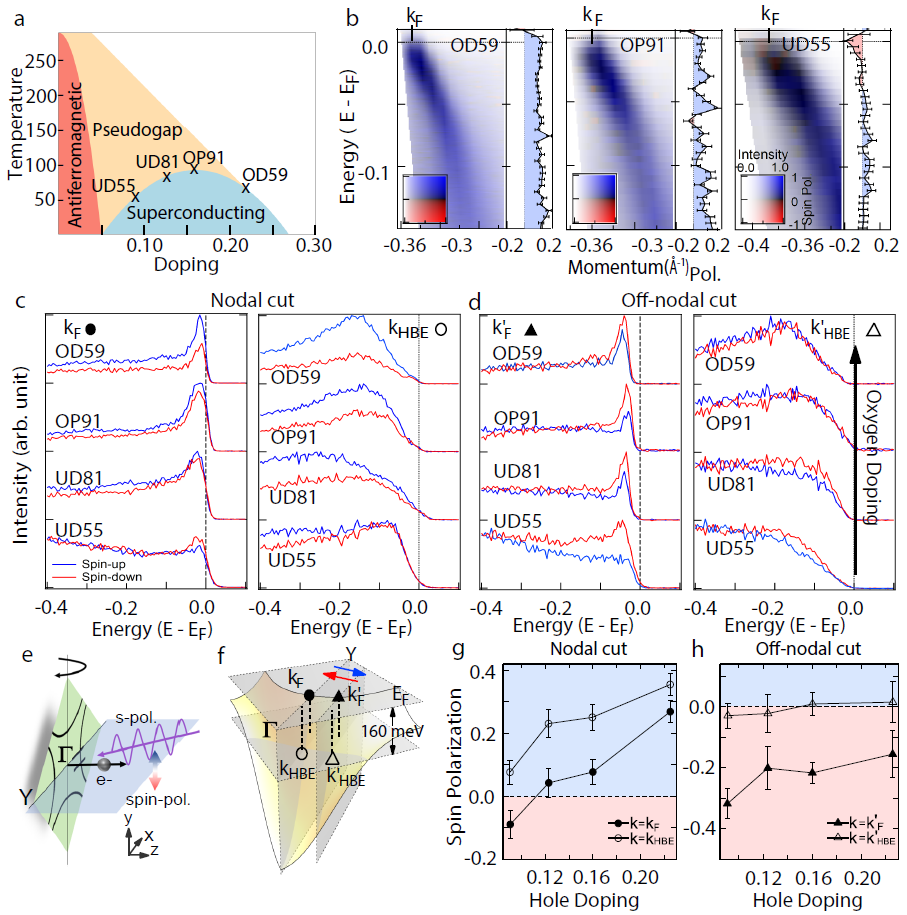}
\caption{\label{fig:fig1} \textbf{Doping dependent spin measurements of Bi2212.}  \textbf{a)} Sketch of the cuprate phase diagram. \textbf{b)} Spin-resolved nodal maps of OD59, OP91, and UD55. The color codes the spin polarization as shown in the inset. To the right of each map is the spin polarization as a function of binding energy at the Fermi momentum $k\rm_F$ corresponding to the nodal point, where $P = (I_{\uparrow}-I_{\downarrow})/(I_{\uparrow}+I_{\downarrow})$. All samples were measured at $\sim$30 K, with the exception of the UD55 sample, which was measured at $\sim$90K. \textbf{c)}, \textbf{d)} Doping dependent spin-resolved EDCs acquired along the \textbf{c)} nodal and \textbf{d)} $\sim$20 degrees from the nodal direction. All cuts are normalized to the OP sample quasiparticle peaks. \textbf{e)} Experimental geometry for the laser-based ($h\nu=6$ eV) spin ARPES experiments. For details see the methods section. \textbf{f)} Location of measured momenta in energy-momentum space, where ($k\rm_F$,$k'\rm_F$) are the locations of the coherent peaks and ($k\rm_{HBE}$, $k'\rm_{HBE}$) are those of the incoherent humps. \textbf{g)}, \textbf{h)} Doping dependence of the total spin polarization in the nodal and off-nodal direction. The error bars are extracted as $\Delta P=1/S\sqrt{N}$ where $S$ is the Sherman function and $N$ is the photoemission intensity\cite{Kessler1976}.}
\end{figure*}

{\bf Doping dependent spin measurements of double layer Bi2212.} Figure~\ref{fig:fig1} shows the spin-resolved ARPES spectra for double layer Bi2212 samples at four different dopings with the respective location on the phase diagram indicated by cross in Fig. ~\ref{fig:fig1}a: underdoped ($x$ = 0.091,  $T\rm_C$ = 55K (UD55), and $x$ = 0.124,  $T\rm_C$ = 81K (UD81)), optimally doped ($x$ = 0.16,  $T\rm_C$ = 91K (OP91)), and overdoped ($x$ = 0.226,  $T\rm_C$ = 59K (OD59)), as shown in Fig. ~\ref{fig:fig1}b-d. Data were collected with {\it s}\,-\,polarized light (Fig. ~\ref{fig:fig1}e) with the measured spin component in-plane and perpendicular to the nodal direction ($\Gamma$ - Y), as schematically shown in Fig.~\ref{fig:fig1}f. Fig.~\ref{fig:fig1}b shows ($E\; vs\; k_{\rm\Gamma Y}$) spin-resolved maps for some representative doping (UD55, OP91, and OD59), where the color represents the direction of the spin polarization along the Fermi surface - blue is positive or ``spin-up" and red is negative or ``spin-down" as defined by the arrows in Fig.~\ref{fig:fig1}f. The spin-resolved nodal maps are mostly blue, suggesting that the spin-up channel is dominating. To the right of each map we show the extracted spin polarization as a function of binding energy in correspondence of the Fermi vector, which reveals a clear overall decrease in magnitude from the overdoped sample to the underdoped ones.

To study in detail the doping dependence of the observed spin asymmetry, in Fig.~\ref{fig:fig1}c,d we show the nodal and off-nodal energy distribution curves (EDCs) for the coherent quasiparticle peak at the Fermi vector ($k\rm_F$) and the incoherent peak at momentum $k\rm_{HBE}$, corresponding to binding energy $E\rm_B = 160$ meV. Two main trends are observed in the data: the first is a decrease of the spin polarization from overdoped to underdoped samples for both coherent and incoherent quasiparticles; the second is the shift of spin polarization from positive to negative as a function of momentum, going away from the nodal direction (compare Fig. ~\ref{fig:fig1}c and Fig. ~\ref{fig:fig1}d).
For the quasiparticle peak at the node, the positive spin polarization present in the overdoped sample decreases with less oxygen doping and eventually reverses to become negative for the heavily underdoped sample. The same negative trend persists for the incoherent part of the spectra, with an overall decrease of $\sim$25-30\% for both momenta. Away from the nodal direction (Fig.~\ref{fig:fig1}d), a decrease in the spin polarization is still observed, but for both the coherent and the incoherent part of the spectra the total decrease is a lot smaller ($\sim$3-13\%). 
The doping dependence of the spin polarization clearly suggests it to be an intrinsic property of the material, rather than an extrinsic effect due to a phase difference in the photoemission process \cite{Fanciulli2017a}, and a manifestation of a bulk property, as will be discussed in more details later.

In addition to the reversal of the direction of spin polarization as a function of doping and momentum, the overall magnitude of the polarization also shows an interesting doping dependence. Along the nodal direction (Fig.~\ref{fig:fig1}c), the spin-up channel shows stronger intensity with respect to the spin-down channel at both $k\rm_F$ and $k\rm_{HBE}$ for all dopings except for the underdoped sample.
For the quasiparticle peak, the relative difference of the two spin channels decreases for both momenta and eventually reverses as the hole doping reduces. Fig.~\ref{fig:fig1}d shows that closer to the antinodal region, the quasiparticle peaks ($k'\rm_F$) for all dopings have stronger negative spin intensity, with the most underdoped crystal showing the strongest difference.
A similar pattern in the spin polarization can also be found at higher binding energy ($k'\rm_{HBE}$). 
These trends are summarized in Fig. ~\ref{fig:fig1}g and ~\ref{fig:fig1}h, where we show the spin polarization of the quasiparticle peaks and incoherent humps as a function of doping. The data show a clear overall increase of the spin polarization from negative to positive as the oxygen doping level increases for both nodal and off-nodal directions, while the magnitude follows an opposite trend between the two directions.

\begin{figure*}\includegraphics[width=16 cm]{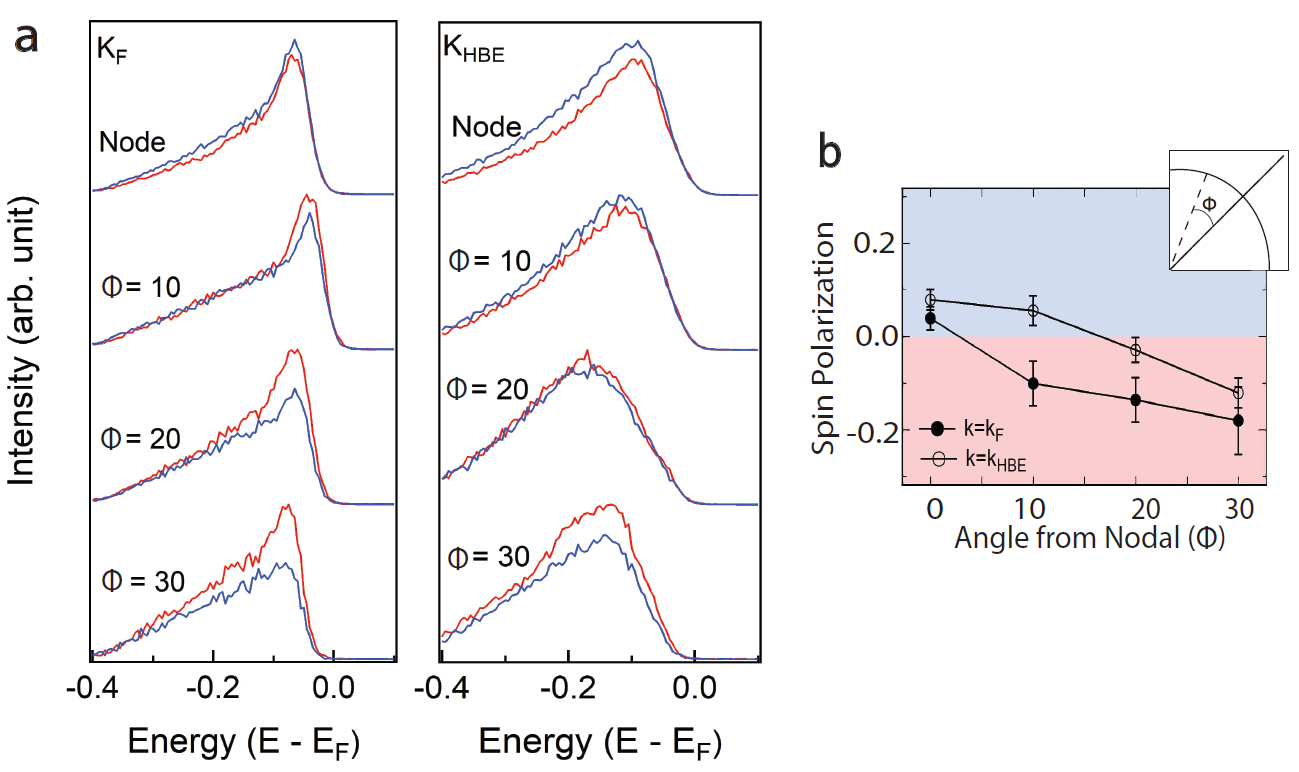}
\caption{\label{fig:fig2}\textbf{Spin polarization of single layer Bi2201} \textbf{a)} Spin-resolved spectra of single layer Bi2201 ($T\rm_C$ = 35 K) at the Fermi level (left) and at the incoherent hump (right) measured at 30\,K. The spin component is in-plane and tangential to the Fermi surface (normal to $\Gamma$-Y), in one quadrant shown in the inset of (b). \textbf{b)} Spin polarization at $k\rm_F$ (quasiparticle peak) and $k\rm_{HBE}$ (incoherent hump) along the Fermi surface.}
\end{figure*}

 {\bf Spin polarization of single layer Bi2201.} To demonstrate that the presence of non zero spin texture extends also to other members of the Bi-family, and is not a property only of the double layer Bi2212, in Figure ~\ref{fig:fig2} we study the momentum dependence of the spin dependent electronic structure for single layer Bi2201. The spin resolved EDCs are shown as a function of the azimuthal angle from the nodal direction both at $k\rm_F$ and the incoherent momentum. Similarly to the double-layer case, we observe a non-zero spin polarization along the nodal direction, with an overall decrease of the positive spin component moving away from it, for both the coherent and incoherent peaks (Fig.~\ref{fig:fig2}a), as summarized in Fig.~\ref{fig:fig2}b, where the spin polarization is reported as a function of momentum, for the two momenta. The overall spin polarization resembled the one discussed in Figure \ref{fig:fig1} for double layer Bi2212, pointing to a universal origin of the observed effect.

\begin{figure*}\includegraphics[width=16 cm]{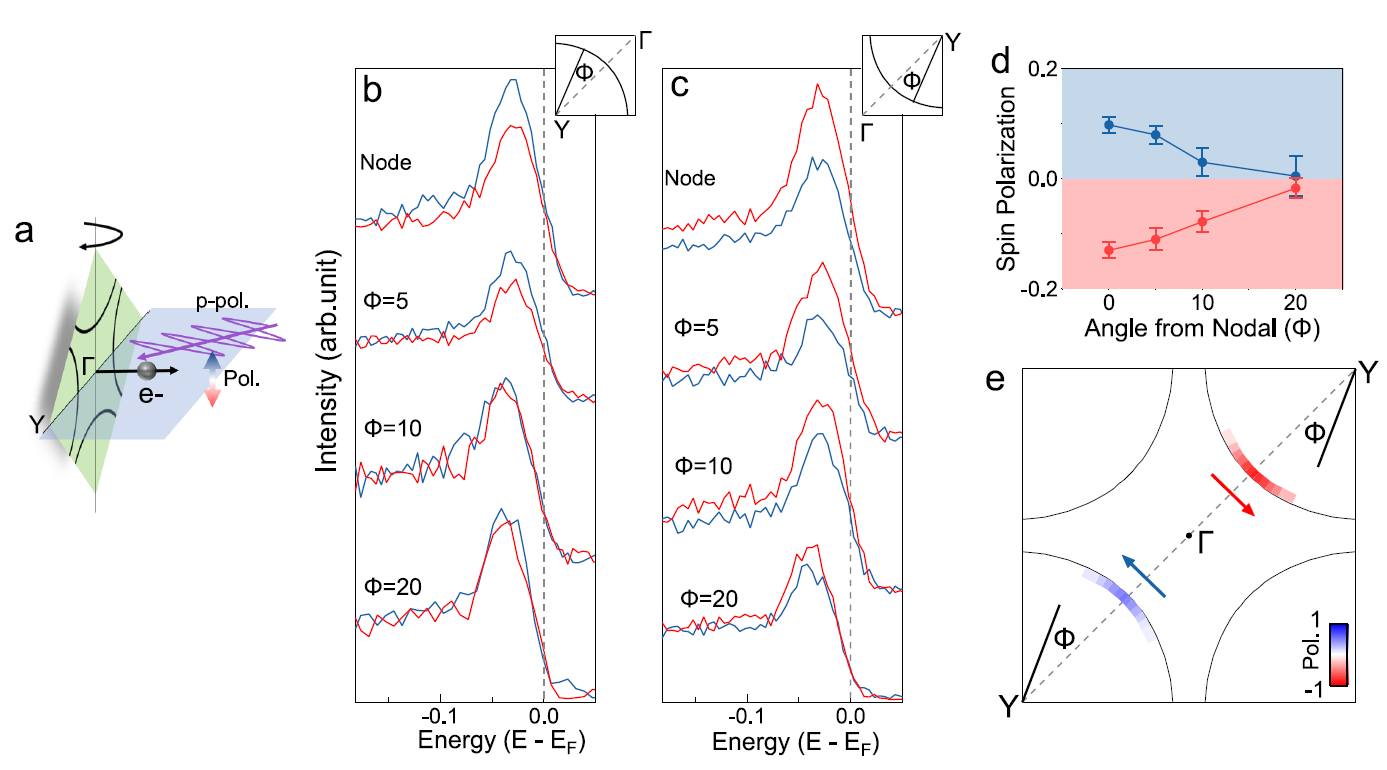}
\caption{\label{fig:fig3} \textbf{In-plane spin polarization along two opposite quadrants of Fermi surface of UD81K Bi2212 measured at 25\,eV.}  \textbf{a)} Experimental geometry for the synchrotron measurements. \textbf{b)} The spin-resolved spectra measured at 10\,K corresponding the spin component perpendicular to the $\Gamma$-Y direction measured along the left-bottom quadrant of Fermi surface as plotted in (e). The momenta along the Fermi surface where we obtain the EDCs are marked by the angle $\phi$. \textbf{c)} Same with (b) but acquired at the right-top branch of the Fermi surface. The spin components perpendicular to the $\Gamma$-Y direction show the opposite spin signs across the $\Gamma$ point. \textbf{d)} Spin polarization at the Fermi vectors as a function of angle $\phi$ from the nodal direction along the Fermi arc.}
\end{figure*}

{\bf Pattern of spin polarization across momentum space.} To better understand the $\it{k}$-dependent pattern of spin polarization across momentum space, in Figure ~\ref{fig:fig3} we show the evolution of the spin polarization of UD81K Bi2212 across the Brillouin zone center $\Gamma$ point, within the first Brillouin zone. Data were collected with {\it p}\,-\,polarized light (Fig. ~\ref{fig:fig3}a) and $h\nu$\,=\,25\,eV photons. Spin resolved EDCs are shown in Fig.~\ref{fig:fig3}b and Fig.~\ref{fig:fig3}c for different momenta near the node, and along the two opposite quadrants of Fermi surface across $\Gamma$ point. The corresponding spin polarization are displayed in Supplementary Fig. S4.
At the node of the quadrant shown in the inset of Fig.~\ref{fig:fig3}b, the intensity is higher for the spin up component, leading to a positive spin polarization (Fig.~\ref{fig:fig3}d). The polarization decreases as we move away from the node and eventually vanishes for $\it{k}$ approaching approximately 1/3 of the Brillouin zone ($\phi$ =20). Although a similar behavior is observed for the spin polarized EDCs in the opposite quadrant (Fig.~\ref{fig:fig3}c), with the absolute value of the polarization decreasing as we move away from the node, the overall spin polarization is reversed in this quadrant (Fig.~\ref{fig:fig3}d). This can be nicely observed also in the spin polarized Fermi surface, shown in Fig.~\ref{fig:fig3}e, where the spin polarized intensity at the Fermi momentum is plotted. This observation is reminiscent of a similar behavior reported for OD58K sample \cite{Gotlieb2018}. We note that the following observation might appear in contrast with previous report\cite{Iwasawa2023} where the overall magnitude of the spin polarization is by a factor of $\sim$4 smaller and no reversal of the spin polarization across the $\Gamma$ point is observed. These differences can be easily accounted for by matrix elements effects, sample quality and alignment (see Supplementary Note for a detailed description). Overall, the reversal of the spin component across the Brillouin zone center points to a spin polarization that not only is a function of $\it{k}$, but also respects time reversal symmetry by switching sign across the $\Gamma$ point.

\begin{figure*}\includegraphics[width=16 cm]{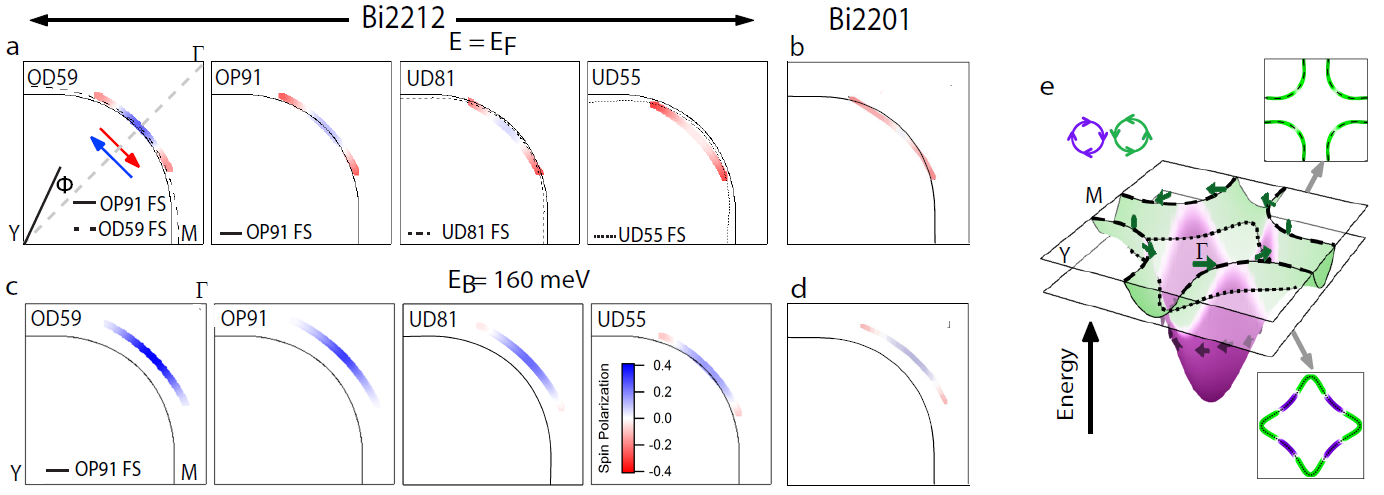}
\caption{\label{fig:fig4} \textbf{Doping dependence of spin-polarized constant energy maps of Bi2212 and Bi2201}  \textbf{a)} Spin-polarized Fermi surface maps for the in-plane spin component tangential to the Fermi surface (normal to $\Gamma$-Y), in one quadrant, for Bi2212 samples. \textbf{c)} Spin-polarized high binding energy maps for Bi2212 samples. The solid black line in each panel represents the tight-binding Fermi surface at optimal doping, while the dashed lines are the calculated Fermi surface for overdoped to underdoped, as specified in each panel.  \textbf{b),d)} Spin-polarized Fermi surface map (b) and high binding energy map (d) for the in-plane spin component, in one quadrant, for Bi2201, summarized according to the data in Fig.~\ref{fig:fig2}. \textbf{e)} Sketch of the proposed three-dimensional (3D) spin texture based on the continuous evolution as a function of doping between the polarization shown for OD59 and UD55 in \textbf{a)} and \textbf{c)} and their location in momentum space. The calculated band structure is color-coded with respect to different spin helicity, with green representing counterclockwise helicity around ${\Gamma}$, and purple representing clockwise case. The effect of changing the doping in a Bi2212 crystal is equivalent to taking a horizontal slice of the 3D spin texture at different binding energies, so higher doping will yield spin with more clockwise helicity, and vice versa when decreasing the doping. The data used for producing the 3D spin texture are shown in Supplementary Fig. S5 and Fig. S6.}
\end{figure*}

{\bf Doping-dependence of the full spin texture of Bi2212 and Bi2201.} In Figure ~\ref{fig:fig4} we summarize all the findings and report the doping dependence of the spin-polarized constant energy maps over one quadrant of the first Brillouin zone, at both $E\rm_F$ (Fermi surface, Fig.~\ref{fig:fig4}a,b) and at $E\rm_B$\,=\,160 meV\,(Fig.~\ref{fig:fig4}c,d) for the double layer and single layer Bi-compounds. Fig.~\ref{fig:fig4}b,d are summarized based on the raw data in Fig.~\ref{fig:fig2}. In each case the data reveal an interesting evolution of the spin texture with both momentum and energy.
Calculated Fermi surfaces based on a tight-binding model \cite{RMarkiewicz2005PRB} are also shown (dashed lines) for each doping level together with the calculated optimal doping Fermi level for comparison (solid lines).
Alongside the standard decrease of the hole-like Fermi surface with decreasing doping, we observe a gradual evolution of the spin polarization. Specifically, in the overdoped regime the polarization is positive along the Fermi arc, and becomes negative as we move along the surface toward the antinode.
The constant energy maps at high binding energy are mainly positive in the overdoped and optimally doped regime and gain some negative spin polarization in the underdoped regime toward the edge of the Fermi arc, reminiscent of the Fermi surface spin texture observed for higher doping (compare OD 59K Fermi surface map in Fig.~\ref{fig:fig4}a with UD 55K high energy map in Fig.~\ref{fig:fig4}c).
These similarities, though from different spectral features in the two samples (coherent peak $\it{vs}$ incoherent hump), occur at momenta that are very close to each other as can be seen by the location of the constant energy maps in relation to the OP Fermi surface (solid line). 
Therefore, the apparent complexity of the measured spin as a function of doping and binding energy seems to hide a qualitative universality in the spin texture, shared by all doping levels, where the observed spin polarization depends mainly on the momentum and energy of the photoelectron or in other words on the position of the chemical potential.
Lowering the hole concentration, i.e., going from the overdoped to the underdoped samples, raises the chemical potential, and therefore changing the doping level is phenomenologically equivalent to taking horizontal slices at different binding energies of a universal three-dimensional spin texture.
This is summarized in Fig.~\ref{fig:fig4}e, where we schematically represent the 3D electronic band structures with the corresponding spin polarization, highlighting the presence of two different helicities - counterclockwise (green) and clockwise (purple) around ${\Gamma}$.

\begin{figure*}\includegraphics[width=16 cm]{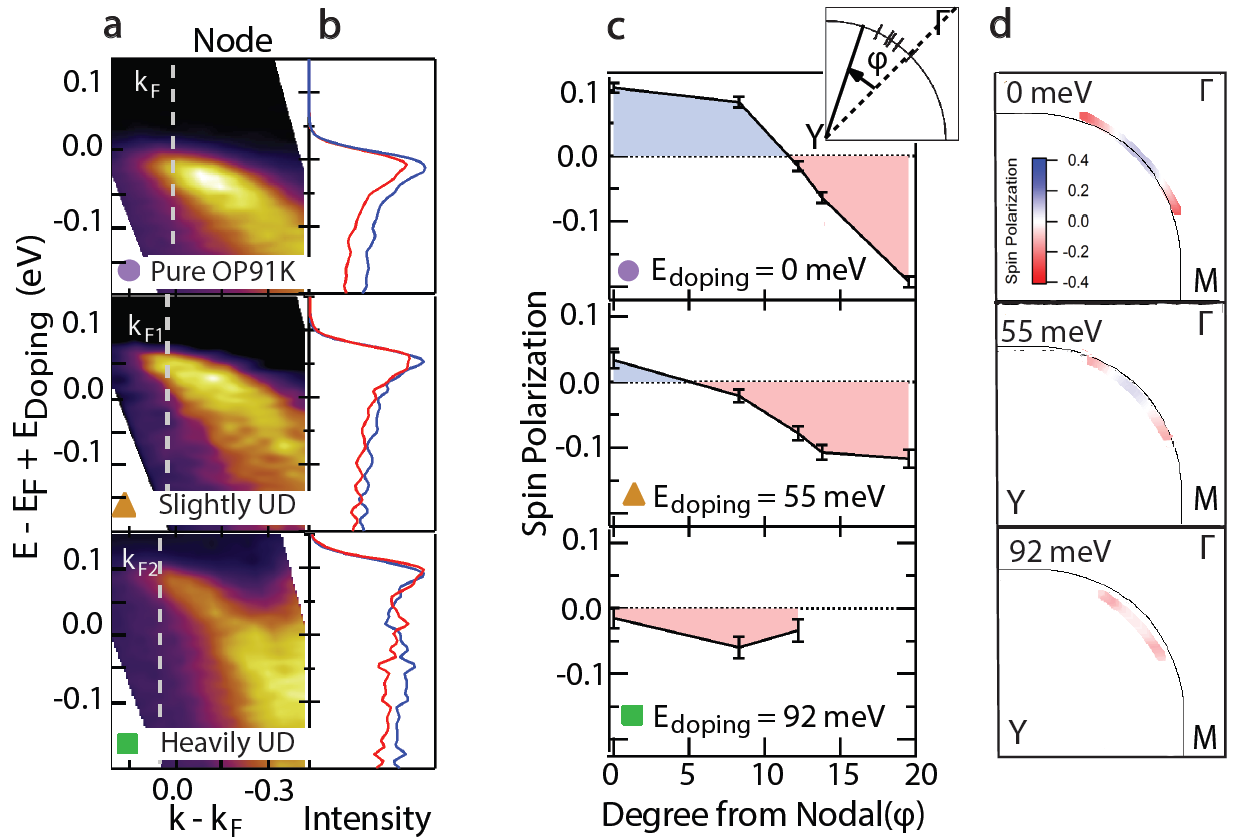}
\caption{\label{fig:fig5} \textbf{Potassium deposition on optimally doped Bi2212.} \textbf{a)} Spin-integrated nodal maps for different levels of K deposition measured at 30\,K. All three maps are adjusted by the Fermi vector of the pure OP91 sample ($k\rm_F$) for ease of comparison. $k\rm_{F1}$ and $k\rm_{F2}$ are the Fermi vectors for the slightly UD and heavily UD samples, respectively. In addition, the energy axis is plotted as $E$\,-\,$E\rm_F$\,+\,$E\rm_{doping}$, where $E\rm_{doping}$ is the energy shift of the bands due to the added electrons, again for ease of visualization of the effects of doping. \textbf{b)} Spin-resolved EDCs at $k\rm_F$. $E\rm_F$ was raised by $\sim$55 meV and $\sim$92 meV after the first and second deposition treatment, respectively. The noise level for each doping is comparable to the bulk-doped counterparts when they are compared within the same energy scale. \textbf{c)} Spin polarization at the Fermi vectors as a function of angle $\phi$ from the nodal direction (see inset) along the Fermi arc. (Raw EDCs shown in Supplementary Fig. S7) \textbf{d)} Constant energy maps at the Fermi surface for comparison with Fig.~\ref{fig:fig4}a.} 
\end{figure*}

{\bf Evolution of the spin polarization upon changing the chemical potential.} The data reported so far demonstrate the existence of an intriguing doping dependent spin-momentum locking across different families of Bi-based compounds. Although the observation of the spin texture under different geometries, photon energies, doping and families, suggests that the observed effect is not extrinsic, to undoubtedly validate its intrinsic bulk nature and the intriguing doping dependence, in Figure~\ref{fig:fig5} we monitor the evolution of the spin polarization upon changing the chemical potential on the same sample. This is achieved via \textit{in situ} doping through deposition of potassium (K) atoms on the cleaved surface of an optimally doped Bi2212 sample. Such an \textit{in situ} doping method has been successfully used in ARPES to study the doping dependence of a variety of materials, such as 2D systems \cite{Zhou2008}, Rashba materials \cite{Crepaldi2012} and cuprate superconductors \cite{Hossain2008}.
The K atoms donate electrons due to the low ionization potential, causing the exposed surface to lose holes (i.e. to become more underdoped), which has practically an effect opposite to that of oxygen doping.
Fig.~\ref{fig:fig5}a shows the spin-integrated ($E\: vs\: k_{\Gamma Y}$) maps as a function of K doping. Clear rises in the chemical potential by 55 meV after the first K doping treatment and by 92 meV after the second are observed, as expected from the evolution of doping from optimal toward underdoped. The evolution of the corresponding spin-resolved EDC spectra are shown in Fig.~\ref{fig:fig5}b. The data reveal a decrease in the spin-up intensity going toward the underdoped sample, similar to the doping dependence reported in Fig.~\ref{fig:fig1}, where the doping change was achieved instead by varying the oxygen content.
In Fig.~\ref{fig:fig5}c the full momentum dependence of the spin polarization is reported, showing that as the K content increases (so the sample becomes more underdoped), the spin polarization at the node becomes more negative, with a decrease from $\sim$11\% to 5\% and then to -2\% after the first and second deposition, respectively.
By showing Fermi surface diagrams (Fig. ~\ref{fig:fig5}d) similar to those in Figure~\ref{fig:fig4}a, we can compare the similarities between the spin textures at similar momenta. The qualitative equivalence with the bulk oxygen doping dependence reported there supports the intrinsic and universal nature of the observed effect in the Bi-based bilayer family, as well as its bulk origin.

{\bf Suppression of spin polarization in lead-doped Bi2212.} The results presented here add two important pieces of information with respect to the previous work \cite{Gotlieb2018}. The first one is the persistence of the spin asymmetry over the entire superconducting dome. The second one is the need for an additional field component to explain the asymmetry in the single layer case. Indeed, previous work has argued in favor of an out of plane electric field generated by the lack of local inversion symmetry within two CuO$_2$ planes in the unit cell \cite{Gotlieb2018}.  
However, this explanation does not hold for the single layer Bi2201 where only one CuO$_2$ is in the unit cell and local inversion symmetry is preserved. 

Here we propose that such a field derives from the local structural fluctuations of the CuO octahedra, a common feature of $p$-type cuprates, including La- \cite{Bozin2000, Billinge2002, Bianconi1996}, Bi- \cite{Bianconi1996a, Bianconi1996b, Slezak2008, Saini1998}, Hg- \cite{Lanzara1999}, Tl- \cite{Pelc2022}, and Y- \cite{MustredeLeon1990} based families, which is also responsible for charge inhomogeneity.
These distortions, caused by the misfit strain between the CuO$_2$ plane and the rocksalt layers \cite{Castro2000, Billinge2002, NPoccia2010ABianconi}, lead to the tilting of the CuO$_2$ planes along the (110) and (010) direction, and the formation of local domains with orthorombic and tetragonal structures \cite{Bianconi1996, Billinge1994}. Such rotation are also known to lead to anharmonic incommensurate modulations and inhomogeneous lattice and charge ordering \cite{Bozin2000, Mihailovic1995, Slezak2008}. In each plane only half of all Cu-O-Cu bonds are buckled, so these tilting lead to a local symmetry breaking within an individual plane, while the overall fourfold symmetry and the average structure are preserved \cite{Bianconi1996}. The proposed source of electric field is in the direction perpendicular to the plane and therefore can account for a Rashba term in the spin texture as previously reported \cite{Gotlieb2018}.  In addition it can provide a universal source of the electric field as a similar effect is also present in the case of double layer Bi2212\,\cite{Bianconi1996}.

A direct way to test this proposal is to eliminate the source of the electric field, through the removal of the structural distortion. This can be achieved through the replacement of Pb with Bi atoms. Indeed, the effect of Pb dopant has been widely studied in the literature, and it has been shown to remove the misfit strain and hence suppress the superstructure along the (110) direction in the BiO plane \cite{Gao1988}, which is directly linked to the local structural fluctuations of the CuO$_2$ planes \cite{Bianconi1996, Saini1998, Slezak2008}. Upon partial substitution of Pb$^{2+}$ ions in place of the smaller Bi$^{3+}$ in Bi2212, the amplitude of the superstructure decreases, eventually disappearing at high Pb concentrations \cite{Liu2019}.

\begin{figure*}\includegraphics[width=16 cm]{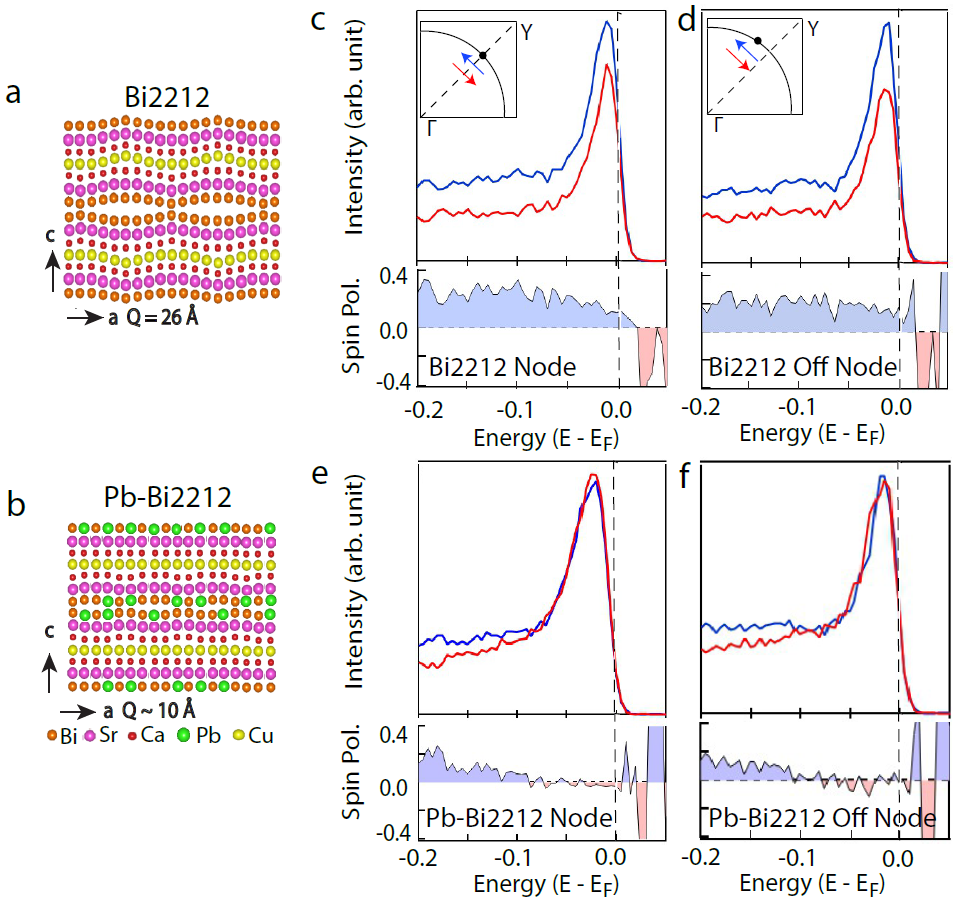}
\caption{\label{fig:fig6}\textbf{Suppression of spin polarization in Pb-doped Bi2212} \textbf{a),b)} Cartoon depiction of the out-of-plane incommensurate distortion in OD Bi2212 (a) and OD Pb-Bi2212 ($x$ = 0.6) (b), respectively. \textbf{c),d)} Spin-resolved EDCs and polarization of the spin component perpendicular to $\Gamma$-Y measured at $k\rm_F$ along the nodal direction (c) and off-nodal direction , $\sim$5 degrees from nodal (d), for OD Bi2212 sample. The momentum positions where these EDCs are taken are marked in the insets of (c) and (d). \textbf{e),f)} Same with (c,d) but the data are taken from OD Pb-Bi2212 ($x$ = 0.6) sample.}
\end{figure*}

This case is studied in Figure ~\ref{fig:fig6}, where we report the comparison between the spin polarization for OD Bi2212 and for Pb-doped Bi2212 (overdoped, $T\rm_C$ = 64 K, $x$ = 0.6). The local lattice fluctuations are strongly reduced, as shown in Fig.~\ref{fig:fig6}a,b. The respective EDC spectra at the node and off-node are shown alongside the spin polarization in Fig.~\ref{fig:fig6}c-f. A striking reduction of the spin polarization is observed in the coherent part of the spectra for the Pb-doped sample with respect to Bi2212, with the imbalance of the spin-up and spin-down intensities that clearly vanishes. This is overall consistent with the proposed model where the spin asymmetry is driven by the local lattice modulation. Another potential scenario to account for the presence of a local symmetry breaking is based on the Internal Quantum Tunneling Polarons scenario, which proposes that the charge imbalance in cuprates can lead to local symmetry breaking, and thus a non-zero local electric field by increasing the Hubbard repulsion term\,\cite{VVelasco2022SConradson}. Furthermore, the suppression of the spin asymmetry appears to occur only in proximity of the quasiparticle peak whereas it is only weakly affected at higher energies, in correspondence of the incoherent part. Whether this is due to some more complex mechanism than the one presented here or is a remnant effect due, for example, to the Pb-induced disorder, is unclear from these data and further studies are needed to determine whether this energy dependence has an intrinsic or extrinsic origin.
\vspace{3mm}

{\bf Conclusions}

In summary, the results presented here hint strongly at an influence of local lattice fluctuations and consequent local symmetry breaking on driving a universal spin texture in hole-doped Bi-based cuprates. Recent theoretical works have proposed that SOC could play an important role in cuprate physics, by stabilizing, for example, the onset of a metallic phase out of an antiferromagnetic insulating background, as recently shown in the case of a Hubbard plaquette \cite{Brosco2018}, or by enhancing the formation of a charge density wave (CDW) and pseudogap phase \cite{Lu2021}, ferromagnetic ordering \cite{Matzdorf2000,Kopp2007}, and even superconductivity \cite{Hurand2015}, in effect turning on different components to shape the symmetry properties of the order parameter \cite{Fischer2011}.
In the underdoped regime, both $\it d$-density wave \cite{Wu2005} and loop-current order \cite{Aji2007} models used to describe the pseudogap phase, rely on a non-zero SOC that couples the orbital current with the spin channel.
In addition, SOC may play a role in the development of a pair density wave state, which could be an alternative and competing ground state to $\it d$-wave superconductivity in cuprates, via local inversion asymmetry \cite{Yoshida2012}.
Moreover, it has been shown that this mutual interplay between SOC and local lattice fluctuations might also have important consequences in defining the shape of the energy landscape, eventually favoring a ``Mexican hat" energy surface over a parabolic one in the limit of strong SOC \cite{Streltsov2020}.
Although many of these predictions require further experimental investigations and the purpose of the present work is not to favor any of them, the results presented here generally demonstrate the importance of including SOC effects in presence of symmetry-breaking lattice distortions in any theories for superconductivity, to examine the role they might play in stabilizing the ground state, and to understand how different phases intertwine and ultimately define the symmetry of the order parameter. For example, how the combination of SOC and lattice symmetry shapes the order parameters would be fascinating to study in connection with the recently discovered superconductivity in 2D heterostructures \cite{Cao2018,Qiu2021}, where lattice-driven local symmetry breaking and its periodicity, strain, and SOC can be tuned via stacking or twisting \cite{Ahn2020,Ernandes2021,SZhao2023PKim, YLee2023NPoccia,MMartini2023NPoccia}. At the same time, this study can inform on new ways to engineer novel spin-orbit coupled superconductors where the imprint of SOC on the order parameter acts as a booster of the superconducting transition \cite{Fischer2011,MMazziotti2021ABianconi,MMazziotti2022AValletta}. Finally, since similar local symmetry breaking due to either a local rotation or tilting of the octahedra structure have been reported in other correlated materials such as pnictides, iridates and ruthenates, where SOC is known to be relevant \cite{Day2018, Cao2018a, Hao2019, Yanase2010, MKim2018AGeorges}, the present work suggests that the interplay between SOC and local lattice fluctuations might play a more universal role in driving the unusual properties of correlated materials and unconventional superconductivity beyond cuprate physics.  
\vspace{8mm}

{\bf Methods}

Single crystal hole-doped Bi2212 (\BSCCOs), Bi2201 (\BSCCOsingles) and lead-doped Bi2212 (\PbBSCCO, $x=0.6$) samples were grown by the floating zone method and the critical temperature of each sample is given in the main text. Previous measurements have shown that the spin remains the same above and below $T\rm_C$ \cite{Gotlieb2018}. All samples were cleaved $in situ$ and kept in ultrahigh vacuum under $\sim$ 5 $\times$ $10^{-11}$ Torr.
From Fig.~\ref{fig:fig1} to Fig.~\ref{fig:fig6} (except for Fig.~\ref{fig:fig3}) and from Supplementary Fig. S5 to Fig. S7, the hole-doped Bi2212, Bi2201 and Pb-doped Bi2212 samples were measured at $\sim$ 30\,K, except for the UD55 sample, which was measured at $\sim$ 90K. The samples were probed with 6 eV photons produced by two stages of second harmonic generation from the output of a Ti:sapphire oscillator. The data were acquired with a time-of-flight spectrometer utilizing an exchange-based spin polarimeter with energy and momentum resolution $\Delta E\approx$ 15 meV and $\Delta k\approx\pm$0.02 \AA$^{-1}$, respectively \cite{Jozwiak2010},
except for UD55, which was taken with a larger aperture yielding momentum resolution $\pm$0.08 \AA$^{-1}$. In Fig.~\ref{fig:fig3} and Supplementary Fig. S1,the data of single crystal UD 81K Bi2212 sample was measured at $\sim$ 10\,K at Beamline 10.0.1.2 of the Advanced Light Source, Lawrence Berkeley National Laboratory (Berkeley, CA).
The quantization axis for the spin measurements is fixed, corresponding to the $y$ axis in Fig.~\ref{fig:fig1}(e) and Fig.~\ref{fig:fig3}(a). The data measured farthest from the nodal point and closest to the antinodal one are measured with a sample tilt of 25$^\circ$ (Fig.~\ref{fig:fig1}(e)) or 6$^\circ$ (Fig.~\ref{fig:fig3}(a)). At these angles there are a maximum factor of $\cos{(25^\circ)}\simeq0.91$ or $\cos{(6^\circ)}\simeq0.99$ between the measured polarization and the actual one in the frame of reference of the sample, i.e., an underestimation of $\sim10\%$ or $\sim1\%$ assuming the polarization vector to be within the surface plane, or smaller otherwise. All the trends shown in Figs.~2-5 are unaffected by this factor.
\vspace{5mm}

{\bf Data Availability}

The data that support the findings of this study are available from the corresponding author upon reasonable request.
\vspace{5mm}

{\bf Acknowledgments}

We thank D. H. Lee, A. Vishwanath, M. Serbyn, and E. Altman for very insightful discussions. This work was primarily supported by the Lawrence Berkeley National Laboratory Ultrafast Materials Sciences program, funded by the U.S. Department of Energy, Office of Science, Office of Basic Energy Sciences, Materials Sciences and Engineering Division, under contract DE-AC02-05CH11231. This research used resources of the Advanced Light Source, which is a DOE Office of Science User Facility under contract DE-AC02-05CH11231.
\vspace{5mm}

{\bf Author Contributions}

H.L., K.C., C.L., K.G., R.M., A.F., and Z.H. carried out the experiments. The single crystal samples were prepared by H.E.'s lab. A.L. was responsible for experimental planning, infrastructure, and experimental design. The manuscript was written by H.L., K.C., and A.L., with input from all other authors.
\vspace{5mm}

{\bf Competing interests}

The authors declare no competing interests.

\end{sloppypar}
\end{document}